\documentclass[12pt]{article}
\usepackage[dvips]{graphicx}

\usepackage{epsfig,amsmath,amssymb,verbatim,mathrsfs}

\def\beq{\begin{eqnarray}}
\def\eeq{\end{eqnarray}}
\def\bea{\begin{eqnarray}}
\def\eea{\end{eqnarray}}

\def\tev{\, {\rm TeV}}
\def\gev{\, {\rm GeV}}

\newcommand{\gsim}{\lower.7ex\hbox{$\;\stackrel{\textstyle>}{\sim}\;$}}
\newcommand{\lsim}{\lower.7ex\hbox{$\;\stackrel{\textstyle<}{\sim}\;$}}

\def\stilde{\widetilde}

\newcommand{\newc}{\newcommand}
\newc{\Nc}{N_{c}}
\newc{\CG}{C_G}
\newc{\gp}{g'}
\newc{\stopi}{\stilde t_i}
\newc{\sboti}{\stilde b_i}
\newc{\staui}{\stilde \tau_i}
\newc{\stopj}{\stilde t_j}
\newc{\sbotj}{\stilde b_j}
\newc{\stauj}{\stilde \tau_j}
\newc{\stopI}{\stilde t_1}
\newc{\stopII}{\stilde t_2}
\newc{\sbotI}{\stilde b_1}
\newc{\sbotII}{\stilde b_2}
\newc{\stauI}{\stilde \tau_1}
\newc{\stauII}{\stilde \tau_2}
\newc{\sstop}{s_{t}}
\newc{\cstop}{c_{t}}
\newc{\ssbot}{s_{b}}
\newc{\csbot}{c_{b}}
\newc{\sstau}{s_{\tau}}
\newc{\cstau}{c_{\tau}}
\newc{\Sstop}{s_{2t}}
\newc{\Cstop}{c_{2t}}
\newc{\Ssbot}{s_{2b}}
\newc{\Csbot}{c_{2b}}
\newc{\Sstau}{s_{2\tau}}
\newc{\Cstau}{c_{2\tau}}
\newc{\salpha}{s_\alpha}
\newc{\calpha}{c_\alpha}
\newc{\Calpha}{c_{2\alpha}}
\newc{\Salpha}{s_{2\alpha}}
\newc{\sbetapm}{s_{\beta_\pm}}
\newc{\cbetapm}{c_{\beta_\pm}}
\newc{\Sbetapm}{s_{2 \beta_\pm}}
\newc{\Cbetapm}{c_{2 \beta_\pm}}
\newc{\sbetaO}{s_{\beta_0}}
\newc{\cbetaO}{c_{\beta_0}}
\newc{\SbetaO}{s_{2 \beta_0}}
\newc{\CbetaO}{c_{2 \beta_0}}
\newc{\vu}{v_u}
\newc{\vd}{v_d}
\newc{\seL}{\stilde e_L}
\newc{\smuL}{\stilde \mu_L}
\newc{\seR}{\stilde e_R}
\newc{\smuR}{\stilde \mu_R}
\newc{\suL}{\stilde u_L}
\newc{\sdL}{\stilde d_L}
\newc{\suR}{\stilde u_R}
\newc{\sdR}{\stilde d_R}
\newc{\scL}{\stilde c_L}
\newc{\ssL}{\stilde s_L}
\newc{\scR}{\stilde c_R}
\newc{\ssR}{\stilde s_R}
\newc{\snue}{\stilde \nu_e}
\newc{\snumu}{\stilde \nu_\mu}
\newc{\snutau}{\stilde \nu_\tau}
\newc{\Gpm}{G^\pm}
\newc{\Hpm}{H^\pm}
\newc{\FFbS}{\overline{FF}S}
\newc{\FFbV}{\overline{FF}V}
\newc{\FSS}{F_{SS}}
\newc{\FSSS}{F_{SSS}}
\newc{\FFFS}{F_{FFS}}
\newc{\FFFbS}{F_{\overline{FF}S}}
\newc{\FSSV}{F_{SSV}}
\newc{\FVS}{F_{VS}}
\newc{\FVVS}{F_{VVS}}
\newc{\FFFV}{F_{FFV}}
\newc{\FFFbV}{F_{\overline{FF}V}}
\newc{\Fgauge}{F_{\rm gauge}}
\newc{\DRbarprime}{$\overline{\rm DR}'$ }
\newc{\DRbar}{$\overline{\rm DR}$ }
\newc{\MSbar}{$\overline{\rm MS}$ }
\newc{\Yu}{{\bf Y}_u}
\newc{\Yd}{{\bf Y}_d}
\newc{\Ye}{{\bf Y}_e}
\newc{\Au}{{\bf a}_u}
\newc{\Ad}{{\bf a}_d}
\newc{\Ae}{{\bf a}_e}
\newc{\bm}{{\bf m}}
\newc{\zhol}{Z^{\rm hol}}

\newc{\rwino}{r_{\tilde W}}
\newc{\rmu}{r_{\tilde H}}
\newc{\ra}{r_A}
\newc{\ccdot}{\!\cdot\!}

\newcommand{\nnmb}{\nonumber}

\newcommand{\met}{\,/\hspace{-0.25cm}E_T}

\newcommand{\lrf}[2]{\left(\frac{#1}{#2}\right)}

\newcommand{\dzero}{D0\!\!\!/}

\begin{document}

\setlength{\baselineskip}{0.2in}


\begin{titlepage}
\noindent
\begin{flushright}
\end{flushright}
\vspace{1cm}

\begin{center}
  \begin{Large}
    \begin{bf}
Higgs Boson Decays to Neutralinos\\
in Low-Scale Gauge Mediation\\
     \end{bf}
  \end{Large}
\end{center}
\vspace{0.2cm}

\begin{center}

\begin{large}
John D. Mason$^a$, David E. Morrissey$^{a,b}$, and David Poland$^a$\\
\end{large}
\vspace{0.3cm}
  \begin{it}
$^{a}$Jefferson Physical Laboratory \\
Department of Physics, Harvard University,\\
17 Oxford Street, Cambridge, MA 02138, USA\vspace{0.5cm}\\
$^{b}$TRIUMF\\
4004 Wesbrook Mall, Vancouver, BC V6T 2A3, Canada\vspace{0.5cm}\\
\end{it}

\end{center}

\center{\today}

\begin{abstract}

  We study the decays of a standard model-like MSSM Higgs boson to pairs
of neutralinos, each of which subsequently decays promptly to a
photon and a gravitino.  Such decays can arise in supersymmetric
scenarios where supersymmetry breaking is mediated to us by gauge
interactions with a relatively light gauge messenger sector
($M_{mess} \lesssim 100\tev$).  This process gives rise to a
collider signal consisting of a pair of photons and missing
energy.  In the present work we investigate the bounds on this
scenario within the minimal supersymmetric standard model from
existing collider data.  We also study the prospects for discovering
the Higgs boson through this decay mode with upcoming
data from the Tevatron and the LHC.

\end{abstract}

\vspace{1cm}

\end{titlepage}

\setcounter{page}{2}


\vfill\eject



\section{Introduction}

  Supersymmetry~(SUSY) is an attractive mechanism to
stabilize the scale of electroweak symmetry breaking against
large quantum corrections from unknown high-energy physics.
However, to be phenomenologically viable SUSY must be broken
at low energies.  One class of models that can achieve this in an
acceptable manner are theories of gauge-mediated supersymmetry
breaking~(GMSB) where supersymmetry breaking occurs in
a \emph{hidden sector}, and is communicated to the \emph{visible sector}
containing the supermultiplets of standard model~(SM) particles
through gauge interactions with a set of \emph{gauge messenger}
particles~\cite{Dine:1981za,Dine:1993yw,Giudice:1998bp}.


  Gauge mediation leads to experimental signatures that are very different
from other forms of SUSY breaking, such as gravity and anomaly mediation,
when the gauge messengers are very light relative to the Planck scale.
In this case, the lightest superpartner particle~(LSP) is generally
the gravitino, and is stable in the presence of $R$-parity
(which we assume in the present work).\footnote{$R$-parity is one simple
way to forbid dangerous operators that can lead to rapid proton decay.}
The lightest standard model superpartner will then decay to
a gravitino and one or more SM states.
For example, if the lightest SM superpartner is a mostly-Bino neutralino,
it can decay to a photon and gravitino.  Such decays can be
prompt on collider time scales for gauge messenger masses below
about $100\,\tev$.  The distinctive photon-rich collider signatures of
this scenario have been studied in a number of works such as
Refs.~\cite{Dimopoulos:1996vz,
Ambrosanio:1996zr,Dimopoulos:1996va,Bagger:1996bt,Lopez:1996gd,Baer:1996hx,
Ambrosanio:1997rv,Kawagoe:2003jv,Hamaguchi:2008hy,Shirai:2009kn}.

  These previous studies of low-scale gauge mediation focused mainly
on a minimal class of GMSB models.  However, more recent investigations
of GMSB scenarios have illustrated that a much broader range of
superpartner spectra are possible within more general
realizations of this mechanism~\cite{Martin:1996zb,Cheung:2007es,
Meade:2008wd,Carpenter:2008wi,Carpenter:2008rj,Carpenter:2008he}.
With the Tevatron running and the LHC about to start up,
it is therefore interesting to consider new and unusual collider
signatures that can emerge in this context~\cite{Berger:2008cq}.
In the present work we investigate a novel Higgs boson
signature within the minimal supersymmetric standard model~(MSSM)
that can potentially arise in generalized GMSB models,
but is impossible in minimal GMSB scenarios.

  The signature that we discuss arises from
the decays of a SM-like Higgs boson to pairs of the lightest neutralinos.
Each neutralino is assumed to subsequently decay promptly
to a photon and a stable gravitino, $\chi_1^0\to \gamma\tilde{G}$,
giving rise to a collider signature
consisting of two photons and missing energy~($\met$).
This decay mode does not occur in standard scenarios of gravity-mediated
supersymmetry breaking because in that case the lightest neutralino is 
(meta-)stable and can only produce an invisible Higgs
signature~\cite{Eboli:2000ze,Belanger:2001am,Davoudiasl:2004aj}.
This decay channel is also impossible in minimal MSSM GMSB models
since the lightest neutralino is constrained to be heavier than
half the mass of the SM-like Higgs boson.  However, Higgs boson
decays to pairs of unstable neutralinos can potentially arise
in generalized GMSB constructions.
To the best of our knowledge, this decay mode has not
been studied previously, although related work has treated the decays
of the heavier non-SM-like Higgs bosons through this
channel~\cite{DiazCruz:2003bx}.
Let us also mention that decays of Higgs bosons directly to a gravitino
and a neutralino have been investigated in Ref.~\cite{Djouadi:1997gw},
while Higgs decays to a heavier and a lighter neutralino
were studied in Ref.~\cite{Chang:2007de}.  Non-standard Higgs decays
in generalized GMSB scenarios within singlet extensions of the
MSSM have also been discussed in Refs.~\cite{Ellwanger:2008py,
Morrissey:2008gm,Mason:2009iq}.

  The layout of this paper is as follows.  In Section~\ref{constr}
we investigate the constraints on this scenario from previous searches
at LEP as well as direct GMSB searches at the Tevatron.
We estimate the prospects of discovering this Higgs boson decay mode
at the Tevatron in Section~\ref{tevatron}.  In Section~\ref{lhc}
we undertake a similar analysis for the LHC.  Finally, Section~\ref{concl}
is reserved for our conclusions.

\section{LEP and Tevatron Bounds on Light Neutralinos\label{constr}}

  We begin by investigating the bounds on light
neutralinos that decay promptly to a photon and a gravitino
from existing particle collider data.  Motivated by recent progress in
non-minimal GMSB models, we consider very general low-energy superpartner
spectra (with a light gravitino) that do not necessarily fit into 
the paradigm of minimal GMSB.  The strongest parameter bounds 
are found to come from
the LEP experiments and the Tevatron searches for direct neutralino
and chargino production in low-scale GMSB.
We also study the maximal size of the branching fraction of the
SM-like Higgs boson to neutralino pairs subject to these constraints.
No dark matter constraints are imposed as the dominant dark matter
component may come from a different sector -- examples include
axions~\cite{Preskill:1982cy,Abbott:1982af} or dark matter in the
supersymmetry-breaking sector~\cite{Dimopoulos:1996gy,Nomura:2005qg,Hamaguchi:2007rb,Mardon:2009gw,Shih:2009he}.
Furthermore, the very light gravitino masses considered here need not
induce any cosmological difficulties~\cite{Viel:2005qj}.

 The Yukawa and gauge couplings of the lightest MSSM neutralino
$\chi_1^0$ to the lightest CP even Higgs boson $h^0$ and the electroweak
gauge bosons are given (in two-component notation) by~\cite{Chung:2003fi,
Dreiner:2008tw}
\bea
-\mathcal{L} &\supset& Y_{h^0\chi^0_1\chi_1^0}\,h^0\chi_1^0\chi_1^0
~+~ Y_{Z^0\chi^0_1\chi_1^0}\,Z^0_{\mu}\,{{\chi}_1^0}^{\dagger}
\bar{\sigma}^{\mu}\chi_1^0\\
&&+~Y_{W^-\chi^0_1\chi_j^+}\,W^-_{\mu}\,{\chi_1^0}^{\dagger}\bar{\sigma}^{\mu}\chi_j^+
~+~ Y_{W^+\chi^0_1\chi_j^-}\,W^+_{\mu}\,{\chi_1^0}^{\dagger}\bar{\sigma}^{\mu}\chi_j^-
~+~ {\rm h.c.},\nnmb
\eea
where
\bea \label{yuk}
 Y_{h^0\chi^0_1\chi_1^0} &=& \left(\cos{\alpha}\,N_{14}^*
+ \sin{\alpha}\,N_{13}^*\right)\left(g'N_{11}^*-gN_{12}^*\right),\\
Y_{Z^0\chi^0_1\chi_1^0} &=& \frac{g}{2\cos{\theta_W}}
\left(|N_{13}|^2 -|N_{14}|^2\right),\nnmb\\
Y_{W^-\chi^0_1\chi_j^+}&=& g\left(\frac{1}{\sqrt{2}}N_{14}V_{j2}^*
-N_{12}V_{j1}^*\right),\nnmb\\
Y_{W^+\chi^0_1\chi_j^-}&=& g\left(\frac{1}{\sqrt{2}}N_{13}U_{j2}^*
+N_{12}U_{j1}^*\right).\nnmb
\eea
 Here the unitary matrix $N$ satisfies the relation $N^*\mathcal{M}^0N^{\dagger}
= {\rm diag}(m_{\chi_1^0},m_{\chi_2^0},m_{\chi_3^0},m_{\chi_4^0})$ where
$\mathcal{M}^0$ is the neutralino mass matrix in the basis
$(\tilde{B}^0, \tilde{W}^3,\tilde{H}_d^0,\tilde{H}_u^0)$.
Similarly, the matrices $U$ and $V$ bi-diagonalize the chargino
mass matrix according to $U^*\mathcal{M}^{\pm}V^{\dagger}$ relative
to the basis $(\tilde{W}^{\pm},\tilde{H}^{\pm})$, and the
angle $\alpha$ describes CP-even Higgs boson mass mixing.
See \cite{Gunion:1984yn} and \cite{Haber:1984rc} for further
discussion and four-component versions of these formulae.

  The results of LEP~I and II and the Tevatron place bounds primarily
on the couplings of neutralinos and charginos to the electroweak gauge bosons.
These constraints imply that in order for the lightest MSSM
neutralino to be light enough to allow superpartner Higgs decays,
it must be mostly Bino since significant Higgsino or Wino components lead to
overly large gauge boson couplings (and unacceptably light charginos).
The Bino, coming from $U(1)_Y$, avoids such couplings due to its
Abelian nature.  This is reflected in the absence of $N_{11}$
in the gauge boson couplings listed in Eq.~\eqref{yuk}.  However,
for the lightest neutralino to couple to the Higgs, it must have
at least some Higgsino content.  From the expressions for Higgs-neutralino
and neutralino-gauge boson couplings, one sees that two Higgsino
(or Wino) mixing factors are required for a mostly Bino state to
couple to a $Z^0$ gauge boson, whereas only one is needed to couple
to the SM-like Higgs boson.  It is this feature that will allow the Higgs
to decay significantly to a Bino-like neutralino in the
MSSM without contradicting LEP constraints.



\subsection{Bounds from LEP and the Tevatron}

  The most constraining LEP searches for a light neutralino decaying
promptly to a photon are those looking for di-photon final states.
LEP~II has searched for di-photons and missing energy up to
$\sqrt{s}\sim 209\,\gev$, which limits $\sigma(e^+e^-\to
\gamma\gamma+\met) \lesssim 10^{-2}\,pb$~\cite{Abdallah:2003np}.
LEP~I running on the $Z^0$ pole also bounds the branching fraction
for $Z^0$ decays to di-photons and missing energy as $BR(Z^0 \to
\gamma\gamma+\met) \lesssim 3\times 10^{-6}$~\cite{Acton:1993kp}.
The $Z^0$ decay bound was derived assuming $Z^0\to \nu\bar{\nu}Z'$
with $Z' \to \gamma\gamma$ and $m_{Z'} = 60\,\gev$.  We expect
the acceptance for $\gamma\gamma+\met$ to be similar for
$Z^0\to \chi_1^0\chi_1^0$ with $\chi_1^0\to \gamma\tilde{G}$,
and we therefore apply the same bound here.
In addition to bounds on the neutralino, the lightest chargino must
typically be heavier than $m_{\chi^+} \gtrsim 103\,\gev$ to be consistent
with the LEP data~\cite{Abdallah:2003xe,Abbiendi:2003sc}.

  Light neutralinos that decay promptly to a photon and a gravitino
are also strongly constrained by Tevatron searches for GMSB.
These searches performed by the CDF and $\dzero$ experiments concentrate
on $\gamma\gamma+\met$ final states.  Their results are presented
in terms of bounds on the specific SPS8 GMSB Snowmass
point~\cite{Allanach:2002nj}, corresponding to minimal gauge
mediation with $\tan\beta=15$, $N_{mess}=1$, $\Lambda = F/M$ a
free parameter, and $M_{mess} = 2\Lambda$.
Given their focus on this specific point, their results do not always
apply to more general scenarios, but we expect their bounds
to carry over to the present case.  The dominant SUSY production
mode for the SPS8 point at the Tevatron is direct electroweak
chargino and neutralino creation. In our numerical scans below,
we will effectively decouple all the superpartners other than the
Bino and the Higgsinos, and electroweak production
of charginos and neutralinos will again be the dominant
Tevatron SUSY production mode.  The primary significant difference
between the spectrum we will study and that of SPS8 is that our lightest
neutralino will be much lighter for a given Higgsino mass.
This may change the decay cascades of the heavier chargino and neutralino
states.  However, since the Tevatron searches concentrate on the
$\gamma\gamma+\met$ signature, the nature of these cascades should
not alter the bounds very much.

  The latest CDF GMSB search makes use of $2.6\,fb^{-1}$ of integrated
luminosity~\cite{cdf-gmsb}.  They demand well-identified photon pairs,
where each photon has $p_T^{\gamma} > 13\,\gev$ and $|\eta| < 1.1$.
In addition, they apply a cut on the event shape to remove jets mis-identified
as photons (MetSig), they demand $H_T = \sum_i\,p_T^i + \met > 200\,\gev$
where the sum runs over all hard photons, leptons, and jets (suitably defined)
to reduce SM backgrounds, and they require
$\Delta\phi(\gamma_1,\gamma_2) < \pi - 0.35$.
After these cuts, the dominant background comes from electroweak
gauge bosons producing genuine $\met$ in the form of neutrinos.
In this signal region, $1.2\pm 0.4$ events are expected while none
are observed, and we interpret this result as an upper bound on the total
chargino and neutralino production cross-section of about
$\sigma_{\chi_{tot}} < 20\,fb$.

  $\dzero$ has a similar but slightly different search based on
$1.1\,fb^{-1}$ of data\;\cite{:2007is}.  They require well-identified
photon pairs such that each photon has $p_T^{\gamma}> 25\,\gev$
and $|\eta| < 1.1$. Using this di-photon sample, they compare the
$\met$ distribution of observed events to the expected background.
Using a likelihood analysis based on the signal shape, they find that
the total chargino plus neutralino production in SPS8 must be less
than about $80\,fb$.  Note that in contrast to the CDF analysis,
they do not impose a cut on $H_T$.

\subsection{Parameter Scans}

  To estimate the maximal size of the Higgs branching fraction
to neutralinos in the MSSM, subject to the bounds from LEP and the
Tevatron on light neutralinos and other superpartners,
we have performed a numerical scan over MSSM parameters
using a modification of NMSSMtools~\cite{Ellwanger:2004xm} and
PYTHIA\,6.4~\cite{Sjostrand:2006za}.  With NMSSMtools we compute the
spectrum and check the LEP bounds on superpartners and the Higgs,
while we use PYTHIA\,6.4 to estimate Tevatron cross-sections.
We have also cross-checked our results obtained in NMSSMtools at the
qualitative level using CPSuperH~\cite{Lee:2003nta}.
In our scans we fix
\bea
M_2 &=& 700\,\gev,~M_3 \,\,=\,\, 800\,\gev,~m_{L_i} \,\,=\,\, 490\,\gev,~
m_{E^c_i} \,\,=\,\, 600\,\gev,
\nonumber\\
m_{Q_i} &=& m_{U_i^c} \,\,=\,\, 2000\,\gev,~
m_{D_i^c} \,\,=\,\, 1970\,\gev,~A_i \,\,=\,\,0\,\gev.
\eea
We also scan over the ranges $M_1 \in [0,80]\,\gev$,
$\mu \in [-500,500]\,\gev$, $\tan\beta \in
[3,15]$, and $m_{A^0} \in [500,2000]\,\gev$.
Since we restrict our attention to spectra with a light Bino
and somewhat heavy scalars, these parameters do not obey the relations
of minimal gauge mediation.  However, they may be realized in more general
GMSB scenarios where the gaugino masses
can be taken as independent parameters~\cite{Martin:1996zb}.
In addition, we have adjusted the scalar masses to satisfy the sum
rules derived in Ref.~\cite{Meade:2008wd}, although deviations away
from them can arise in even more general GMSB scenarios.\footnote{
We do not specify here the origin of the $\mu$ term, but it is important
to keep in mind that mechanisms to generate it may induce
violations of the sum rules in the third generation or
contribute to $A$-terms~\cite{Komargodski:2008ax}.}
The large stop and pseudoscalar Higgs masses ensure that
the lightest Higgs is SM-like and heavier than $114\,\gev$, even for smaller
values of $\tan\beta$.\footnote{Note that the Higgs mass predicted by
NMSSMtools has a theoretical uncertainty of a few GeV.}
We also take somewhat large slepton soft masses
to suppress the $t$-channel selectron contribution to LEP neutralino production.
We have also studied lower stop and Higgs masses, but as we will discuss below,
this does not substantially increase the region of allowed parameters.

\begin{figure}[ttt]
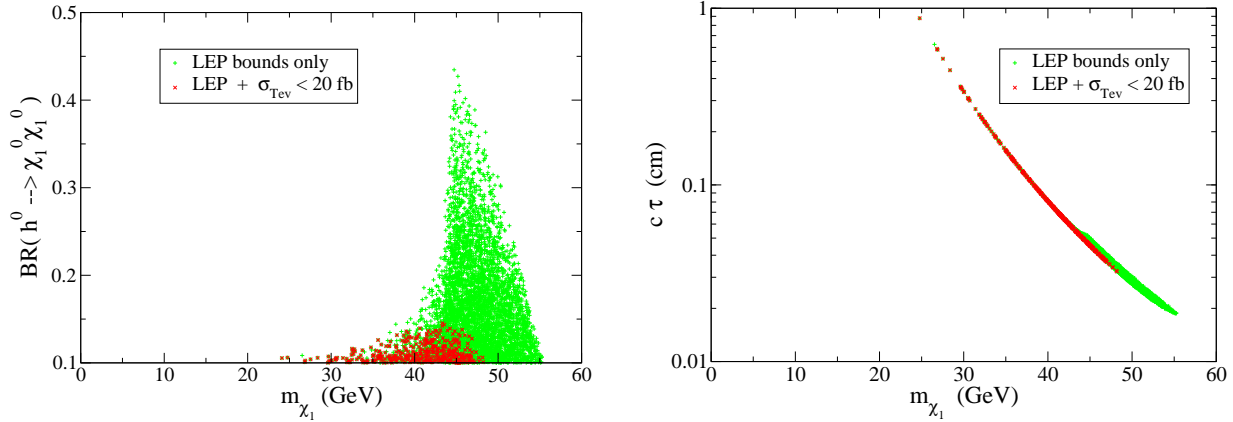

\begin{minipage}[t]{0.47\textwidth}
        \includegraphics[width = \textwidth]{mnbr.eps}
\end{minipage}
\phantom{aa}
\begin{minipage}[t]{0.47\textwidth}
        \includegraphics[width = \textwidth]{msdist.eps}
\end{minipage}
\caption{Higgs boson branching fraction into neutralinos
$h^0\to \chi_1^0\chi_1^0$ (left) and neutralino decay length
to photon plus gravitino (right) as functions of the neutralino mass.
The green points in both plots are consistent with LEP bounds while the
red points are also consistent with the Tevatron search bounds for
neutralinos that decay promptly to photons.  All points in the plot
on the right have $BR(h^0\to\chi_1^0\chi_1^0)>0.1$.}
\label{mnbr}
\end{figure}

  In the left panel of Fig.~\ref{mnbr} we show the values of
the branching fraction $B(h^0\to\chi_1^0\chi_1^0)$ obtained in our
scan, both before and after imposing bounds from the Tevatron
($\sigma_{\chi_{tot}} < 20\,fb$). With only the LEP bounds,
branching fractions as large as $0.45$ are possible for
$m_{\chi_1^0} \gtrsim M_Z/2$.  For lighter masses,
the constraint from $Z^0$ decays limits the Higgsino
content of the light neutralino, and in turn its coupling
to the SM-like Higgs boson.  As a result, the allowed branching
ratio of the Higgs to neutralinos falls rapidly when
$m_{\chi_1^0} < M_Z/2$.  When the additional bounds
from Tevatron GMSB searches are included, the maximal Higgs
branching fraction to neutralinos is reduced to about $0.15$.
Branchings of this size are too small to significantly affect
the LEP-II limits on the SM Higgs mass of about $114\,\gev$
derived from searches for Higgs-strahlung off a $Z^0$ boson with
$h^0\to b\bar{b}$~\cite{Schael:2006cr}.

  In the right panel of Fig.~\ref{mnbr} we show
the (unboosted) decay lengths $c\tau$ of the product neutralinos
as a function of the neutralino mass for a supersymmetry-breaking
scale of $\sqrt{|F|} = 50\,\tev$.  We show only points for
which $BR(h^0\to\chi_1^0\chi_1^0) > 0.1$.  These decay lengths
were obtained using the standard result for minimal gauge
mediation~\cite{Fayet:1979qi,Ambrosanio:1996jn}
\beq
c\tau = 48\pi\,\frac{m_{3/2}^2\,M_{\rm Pl}^2}{m_{\chi_1^0}^5}
\,\frac{1}{|P_{1\gamma}|^2},
\eeq
where $|P_{1\gamma}| = |N_{11}\,c_W+N_{12}\,s_W|$,
and $m_{3/2} = |F|/\sqrt{3}M_{\rm Pl} \simeq 0.6\,\mbox{eV}$.\footnote{
The precise values of $c\tau$ and $m_{3/2}$ will differ in more general
GMSB scenarios, but we expect their values to be qualitatively similar
in many cases.}
This value of $m_{3/2}$ is cosmologically safe~\cite{Viel:2005qj},
but still large enough that direct $Z^0$ and Higgs decays to gravitinos
are negligible~\cite{Djouadi:1997gw,Dicus:1990dy}.  Note that the boosts
here are typically of order unity because, for regions of parameter
space where $B(h^0\to\chi_1^0\chi_1^0) > 0.1$, $m_{\chi^0}$ can not
be significantly lighter than $m_h/2$ due to the bound from $Z^0$ decays.
Among the allowed points shown in Fig.~\ref{mnbr}, we find that
they all have decay lengths less than 2\,cm, about the limit of what
$D0\!\!\!/$ can detect using electromagnetic calorimeter~(ECAL) pointing.
The resolution (based on timing) at CDF is somewhat worse.
In both cases the small decay length before the photon is emitted
is not large enough to interfere with the photon reconstruction
algorithms.  We concentrate on such ``prompt'' photons in the present work,
but it would also be interesting to look at the case of displaced photons.

\begin{figure}[ttt]
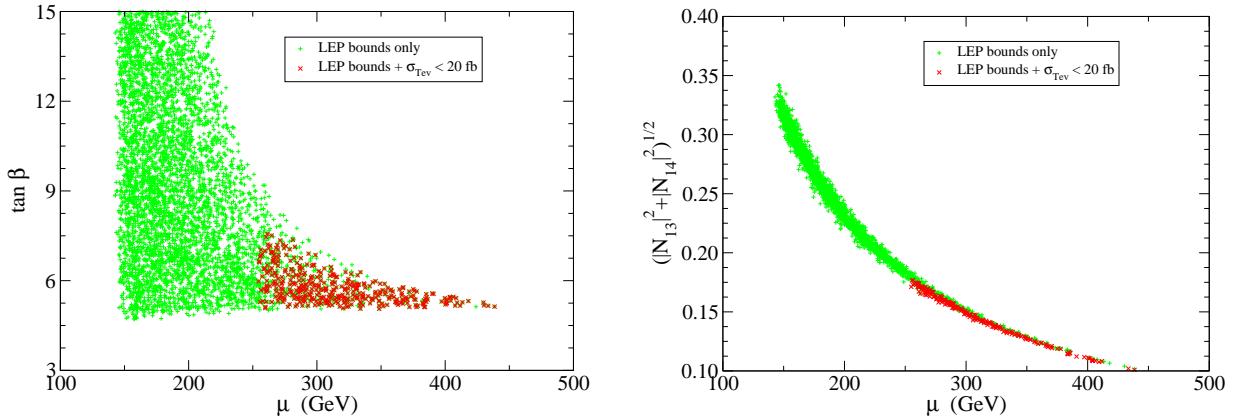

\vspace{1cm}
\begin{minipage}[t]{0.47\textwidth}
        \includegraphics[width = \textwidth]{mspars.eps}
\end{minipage}
\phantom{aa}
\begin{minipage}[t]{0.47\textwidth}
        \includegraphics[width = \textwidth]{hinofrac.eps}
\end{minipage}
\caption{Allowed MSSM parameter regions in the $\mu\!-\!\tan\beta$ plane (left)
and the corresponding Higgsino fraction $(|N_{13}|^2+|N_{14}|^2)^{1/2}$ of
$\chi_1^0$ (right) for which $BR(h^0\to\chi_1^0\chi_1^0) > 0.1$.
The green points satisfy the LEP bounds, while the red points also
avoid the constraints from GMSB searches at the Tevatron.
}
\label{mspars}
\end{figure}

  To see the origin of the reduction in the neutralino branching fraction
when the Tevatron bounds are included, we show in
Fig.~\ref{mspars} the values of $\mu$ and $\tan\beta$ for the
LEP-allowed scan points with $BR(h^0\to \chi_1^0\chi_1^0) > 0.1$
as well as the range of Tevatron production rates.
The constraint imposed on the Higgs
branching ratio restricts the set of points appearing in the
$\mu$-$\tan\beta$ plane.  Tevatron bounds require larger
$\mu$ and $\tan{\beta}$.
However, lower values of $\mu$ and smaller
values of $\tan\beta$ are needed to obtain a significant
Higgs-neutralinos coupling. Let us also point out that our scans
include both positive and negative values of $\mu$, but only
positive values generate a significant Higgs branching fraction,
greater than $0.1$.  This arises because, having fixed $M_1>0$,
positive $\mu$ leads to more Bino-Higgsino mixing.  From
Fig.~\ref{mspars} we also see that the Tevatron direct production
cross-sections of neutralinos and charginos become unacceptably
large for $\mu \lesssim 250\,\gev$.  This is simply the result of
the charginos and heavier neutralinos derived from the Higgsinos
becoming light enough to be produced at the Tevatron without too
much kinematic suppression.

\section{Di-photon Searches at the Tevatron\label{tevatron}}

  To investigate the prospects for discovering a SM-like Higgs boson
at the Tevatron through neutralino decays leading to prompt photons,
we have generated parton-level (but including initial- and final-state
radiation) events in PYTHIA\,6.4~\cite{Sjostrand:2006za} for a
particular MSSM sample point.  The parameter values for this point are
\bea
M_1 \!&=&\! 50\,\gev,~~\mu \,\,=\,\, 300\,\gev,~~\tan\beta \,\,=\,\, 5.5,~~m_{A^0} \,\,=\,\, 1000\,\gev,
\eea
with all other soft parameter taken as in our previous parameter scans.
For these values we find
$BR(h^0\to \chi_1^0\chi_1^0) \simeq 0.11$,
$m_h\simeq 114.7\,\gev$,
$m_{\chi_1^0} \simeq 46.6\,\gev$, as well as a total leading-order
chargino/neutralino Tevatron production cross-section of
$\sigma_{\chi} \simeq 7.2\,fb$.  This point appears to satisfy all
current experimental bounds (not including the DM density).

  Very significant SM backgrounds to the signals we are interested
in come from jets and electrons mis-identified as photons.  We do not
attempt to model this detector-dependent background.
Instead, we simulate signal events while applying cuts and scaling by
efficiency factors to allow for direct comparisons to the backgrounds
tabulated in the most recent published $\dzero$ searches for $h^0\to
\gamma\gamma$~\cite{dzero-hgg} and GMSB~\cite{:2007is}.
Both searches select events containing a pair of photons,
each with $p_T^{\gamma}> 25\,\gev$ and $|\eta| < 1.1$.
In addition, the candidate photons must be isolated and pass a set
of photon identification requirements which differ slightly between
the two analyses.  The $\dzero$ GMSB analysis also considers cuts on
missing transverse energy ($\met$)~\cite{:2007is}.

  The scaling factors we apply consist of a K-factor to account
for higher-order corrections to the cross-section, as well as a
detection efficiency factor $\varepsilon$.  For Higgs production,
we choose the K-factors such that the cross-sections in PYTHIA agree
with published $N^{n}LO$ results.  We consider production through
gluon fusion~(GF)~\cite{Anastasiou:2008tj,deFlorian:2009hc}, $W/Z$-associated
modes~\cite{Ciccolini:2003jy,Brein:2003wg}, and vector boson
fusion~(VBF)~\cite{Aglietti:2006ne}, and we obtain scaling factors
of $K_{GF}:K_{W/Z}:K_{VBF} = 4.7:1.2:1.0$.  The K-factor we use for
chargino/neutralino production at the Tevatron
is $K_{\chi} = 1.1$~\cite{:2007is}.

  The efficiency factors $\varepsilon$ quoted in
Refs.~\cite{dzero-hgg,:2007is} include both the acceptance probability
$\mathcal{A}$ that the signal passes the given cuts, as well as the
reduced efficiency $\tilde{\varepsilon}$ to reconstruct the
fraction of signal events passing the cuts.  We assume that these
factorize according to
\beq
\varepsilon = \mathcal{A}\,\tilde{\varepsilon}.
\eeq
While the acceptance $\mathcal{A}$ clearly depends on the process under study,
we assume further that the reduced efficiency $\tilde{\varepsilon}$
is constant for all events that pass the cuts.
To extract the reduced efficiencies $\tilde{\varepsilon}$, we
estimate the signal acceptances for di-photons from a $120\,\gev$
Higgs boson or from the SPS8 GMSB Snowmass point ($\Lambda = 100\,\tev$)
by simulating these signals and imposing the relevant cuts within
PYTHIA.  We find $\mathcal{A}_{h} \simeq 0.42$ and
$\mathcal{A}_{GMSB} \simeq 0.53$.  Given the $\dzero$ values
$\varepsilon_{h} \simeq 0.20$ for di-photons from a $120\,\gev$
Higgs~\cite{dzero-hgg}, and $\varepsilon_{GMSB} \simeq 0.20$ for
the SPS8 point~\cite{:2007is}, we deduce that $\varepsilon_{h}
\simeq 0.47$ and $\varepsilon_{GMSB} \simeq 0.38$.  Presumably the
difference between these values is due to the slightly different
photon identification requirement used in the two $\dzero$ analyses.

\begin{figure}[ttt]
\vspace{1cm}
\begin{center}
        \includegraphics[width = 0.7\textwidth]{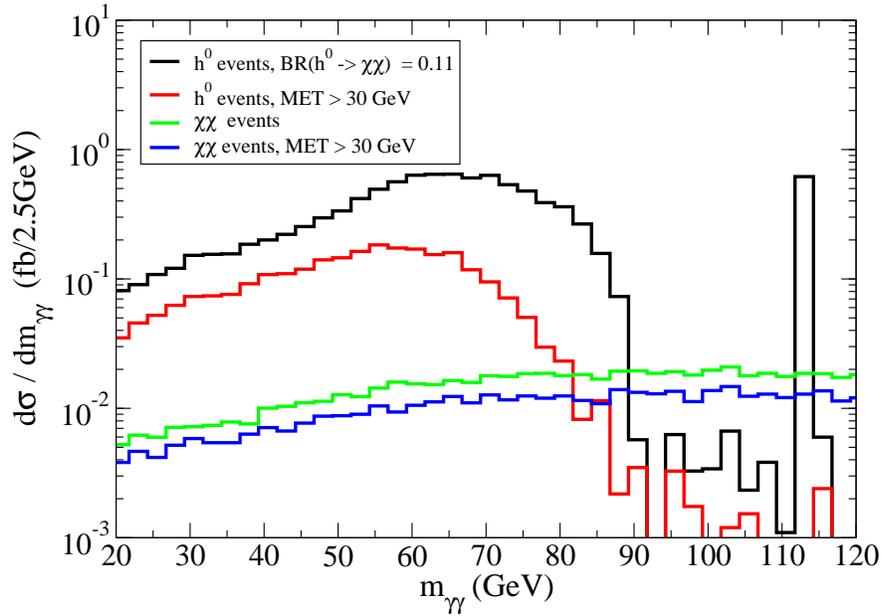}
\end{center}
\caption{Di-photon signal events at the Tevatron from the Higgs
boson and chargino/neutralino cascade decays.  The distributions
without a $\met$ cut are scaled by K-factors and efficiencies to
allow for direct comparison with the backgrounds of
Ref.~\cite{dzero-hgg}, while the distributions with the imposed
cut $\met > 30\,\gev$ are scaled to make them comparable to the
backgrounds in Ref.~\cite{:2007is}.} \label{mgg-tev}
\end{figure}

  In Fig.~\ref{mgg-tev} we show the di-photon invariant mass distribution
of Tevatron events originating from the $h^0$ Higgs, both with
and without a cut on missing energy.  We also show in this figure the
di-photon mass distribution from the endpoints of chargino and neutralino
cascades.  All events are required to have at least two photons with
$p_T^{\gamma}> 25\,\gev$ and $|\eta| < 1.1$.  The
distributions for which a $\met$ cut was not imposed have been
scaled to be directly comparable to the backgrounds tabulated in
the $\dzero$ di-photon Higgs search in Ref.~\cite{dzero-hgg}, while the
distributions with $\met > 30\,\gev$ are scaled to be comparable
with the $\dzero$ GMSB search~\cite{:2007is}.\footnote{Note that
the QCD background in the Higgs search is estimated from a
sideband analysis that cuts out the window $m_h\pm 15\,\gev$.
Di-photons from $h^0\to\chi_1^0\chi_1^0$ will also contribute in
this sideband region.  However, the signal is much smaller than
the background within any one bin, so this effect will be tiny.}
No detector smearing effects have been included in this plot.

   Fig.~\ref{mgg-tev} illustrates that di-photons from the Higgs
dominate strongly over those from chargino and neutralino cascades
in the low $m_{\gamma\gamma}$ region.  Di-photons from Higgs
decays to neutralinos are very spread out relative to direct
di-photon decays, which appear as the sharp spike around
$m_{\gamma\gamma} \simeq 115\,\gev$.  The broad distribution of
di-photon events from Higgs decays to neutralinos is also seen
to fall off sharply near $90\,\gev$.  This arises from a kinematic
endpoint in the invariant mass distribution,
\beq
m_{\gamma\gamma} \leq \frac{2\,m_{\chi_1^0}^2}
{m_{h}-\sqrt{m_{h}^2-4m_{\chi_1^0}^2}}.
\eeq
with equality occurring when both photons are emitted parallel
to the direction of the outgoing neutralinos.  In principle, a measurement
of this endpoint together with an independent determination of the Higgs
boson mass from the location of the diphoton resonance would yield the
neutralino mass.

  Comparing our simulation to the background estimate in Ref.~\cite{dzero-hgg},
the signal from $h^0\to \chi_1^0\chi_1^0$ appears to be too small
to be observable.  From a pure counting perspective, there are about
$S=7$ signal events with $50\,\gev < m_{\gamma\gamma} < 90\,\gev$ relative to
about $B=8000$ expected for background (3 and 2000 for $70\,\gev <
m_{\gamma\gamma} < 90\,\gev$), yielding a significance $S/\sqrt{B}
< 0.08$ ($0.07$) for $1\,fb^{-1}$ of data.\footnote{Our background
estimates are based on the very rough parametrization
$\frac{d\sigma}{dm_{\gamma\gamma}} \simeq
(10\,fb/\gev)(\frac{m_{\gamma\gamma}}{150\,\gev})^{-3.7}$
of the background given in Fig.~3 of Ref.~\cite{dzero-hgg}.}
Increasing the integrated luminosity to $10\,fb^{-1}$ is still not
enough to yield a visible signal.  Even with much more luminosity,
the very small value of $S/B$ in this channel will likely make it
impossible to observe.

  To reduce the SM background, a simple option is to impose a cut on
missing energy $\met$, as considered in the $\dzero$ searches for gauge
mediation with prompt photons~\cite{:2007is}. Adding a
cut of $\met > 30\,\gev$ to the signal detection requirements
described above, we find a reduction in the $h^0\to
\chi_1^0\chi_1^0 \to \gamma\gamma\tilde{G}\tilde{G}$ signal by
more than a factor of 2.  In comparison, the SM background is reduced
by a much larger factor, as can be seen in Fig.~1 of Ref.~\cite{:2007is},
which is based on an integrated luminosity of $1.1\,fb^{-1}$.
For a fiducial integrated luminosity of $1\,fb^{-1}$, we find $2.7$
signal events with $m_{\gamma\gamma} < 100\,\gev$ relative
to an expected background of $9.8\pm 1.0$ (see Table~I of Ref.~\cite{:2007is}),
corresponding to a statistical significance of $S/\sqrt{B} = 0.86$.
Scaling up to an integrated luminosity of $10\,fb^{-1}$, this yields a
statistical significance close to $3$, enough for a $95\%$ exclusion
and nearly enough for preliminary evidence.  This is comparable to or
better than estimates for the combined Tevatron reach for a SM Higgs
boson with mass below $m_h < 130\,\gev$, even with improvements in detection
efficiency~\cite{Draper:2009fh}.

\begin{figure}[ttt]
\vspace{1cm}
\begin{center}
        \includegraphics[width = 0.7\textwidth]{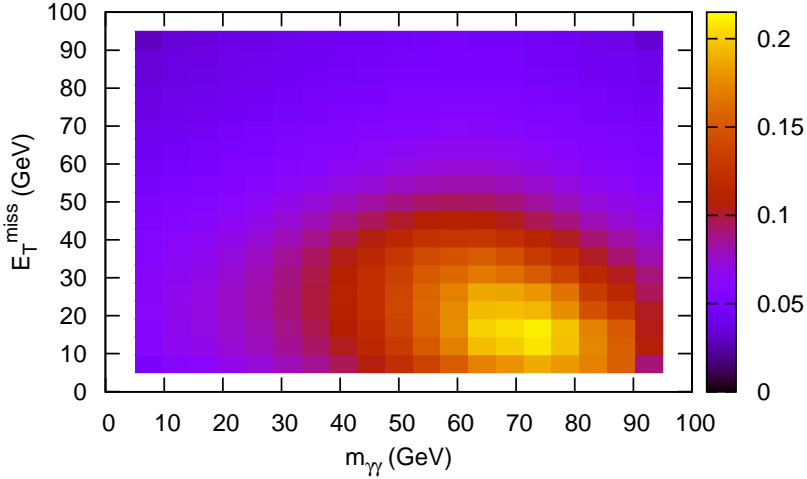}
\end{center}
\caption{Distribution of di-photon events from $h^0\to
\chi_1^0\chi_1^0$ in the $m_{\gamma\gamma}$-$\met$ plane subject
to the $\dzero$ photon requirements discussed in the text
($p_T^{\gamma}>25\,\gev$, $|\eta|< 1.1$, and photon ID). The
$z$-axis units are $fb/(10\,\gev)^2$ after applying the rescalings
relevant for the $\dzero$ GMSB search.}
\label{mgg-et}
\end{figure}

  Our simple significance estimate based on counting will
be degraded once systematic uncertainties and the small number
of total events are factored in.  On the other hand, there is a
great deal of shape information we have not used.  The di-photon
spectrum from Higgs decays to neutralinos is peaked towards
$m_{\gamma\gamma} \sim 70\,\gev$, whereas our expectation is that
the SM backgrounds will fall off quickly with $m_{\gamma\gamma}$.
Therefore modifying the cuts within the $m_{\gamma\gamma}$-$\met$
plane and fitting to a signal shape in this plane using a
two-dimensional log-likelihood analysis has an excellent chance of
improving the signal significance.  This technique would also
reduce the sensitivity to systematics.  We illustrate the
distribution of di-photon events passing the basic photon
identification requirements discussed above
($p_T^{\gamma}>25\,\gev$, $|\eta|< 1.1$, and photon identification) in
Fig.~\ref{mgg-et}.  We should also mention that somewhat relaxing the
requirements on $p_T^{\gamma}$ may also help improve the signal significance.
Our results suggest that this channel would benefit from further study by
the experimental collaborations.  Note that the CDF GMSB search technique
as presented in Ref.~\cite{cdf-gmsb} is less useful for finding
this Higgs mode since their cut of $H_T > 200\,\gev$ removes
nearly all of this signal.

\section{LHC Searches\label{lhc}}

  For a Higgs boson of mass less than $125\,\gev$, the most promising
search mode at the LHC is inclusive production followed by a decay
to di-photons.  In the mass region $115 < m_{h} < 125\,\gev$,
both ATLAS and CMS should be able to discover a SM-like Higgs with
about $15\,fb^{-1}$ of data.  Here, we attempt to extrapolate
these analyses to the broader di-photon spectrum expected from
Higgs decays to unstable neutralino pairs.

  The ATLAS inclusive $h^0\to \gamma\gamma$ analysis presented in
Ref.~\cite{Aad:2009wy} requires two well-reconstructed photons
with $p_T^{\gamma} > 40\,\gev,\,25\,\gev$ and $0<|\eta|<1.37$ or
$1.52< |\eta| < 2.37$.  The total detection efficiency for
di-photons from a $120\,\gev$ Higgs after applying these cuts
(along with photon ID requirements) is $\varepsilon \simeq
0.36$ in the absence of pile-up. As in our Tevatron analyses, we
assume that $\varepsilon = \mathcal{A}\tilde{\varepsilon}$, where
$\mathcal{A}$ is the probability that the signal passes the
$p_T^{\gamma}$ and $\eta$ cuts, and that the reduced efficiency
$\tilde{\epsilon}$ is effectively constant.  Simulating $h^0\to
\gamma\gamma$ events in PYTHIA, we deduce $\tilde{\epsilon} \simeq
0.59$, which is consistent with the general photon detection
efficiencies discussed in Ref.~\cite{Aad:2009wy}.
When simulating signal events from $h^0\to\chi_1^0\chi_1^0$, we
consider GF, $W/Z$-associated production, and VBF, and scale the
production cross-sections by factors of
$K_{GF} = 3.2$~\cite{Anastasiou:2008tj,deFlorian:2009hc},
$K_{W/Z} = 1.3$~\cite{Ciccolini:2003jy,Brein:2003wg},
and $K_{VBF} = 1$~\cite{cms-tdr}, where the particular
values are chosen to rescale the PYTHIA values to higher-order estimates.

\begin{figure}[ttt]
\vspace{1cm}
\begin{center}
        \includegraphics[width = 0.7\textwidth]{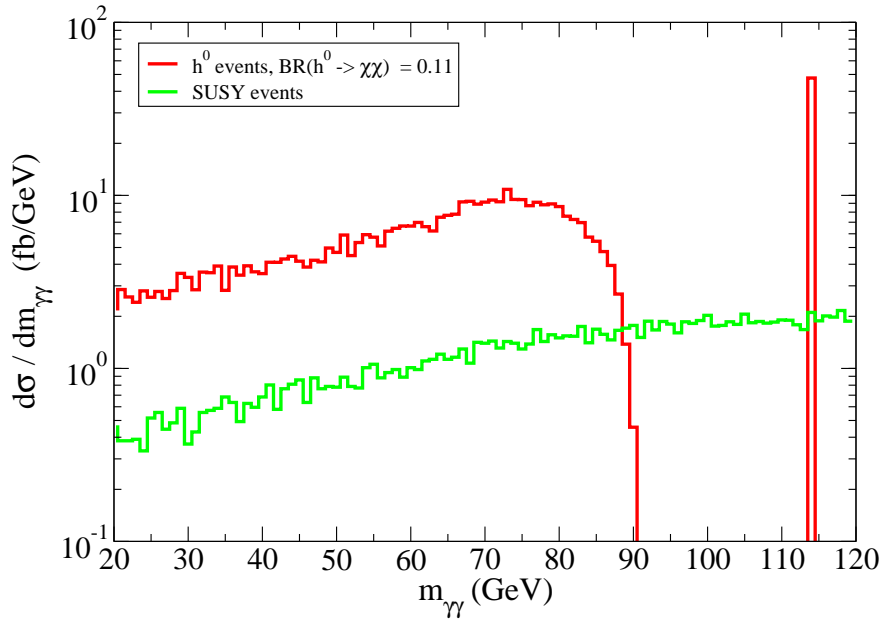}
\end{center}
\caption{Photon invariant mass distributions from Higgs decays and
GMSB SUSY events with cuts and efficiencies as in the ATLAS
di-photon Higgs analysis ($p_T^{\gamma} > 40,\,25\,\gev$, $|\eta|
< 2.37$, and photon identification). } \label{mgg-lhc}
\end{figure}

  In Fig.~\ref{mgg-lhc} we show the expected Higgs boson signal
for the MSSM sample point described in the previous section
after imposing the cuts and scalings discussed above. In this
figure we also show the di-photon signal from SUSY cascade events.
The dominant contribution in this case comes from gluino
production ($M_3 = 800\,\gev$), but electroweak neutralino and
chargino production is also significant.  The SUSY signal with
$m_{\gamma\gamma} < m_h$ is seen to be subleading relative to the
Higgs contribution.  To evaluate the significance of the $h^0\to
\chi_1^0\chi_1^0$ di-photon signal, we estimate the SM background
by making a power-law extrapolation to the inclusive di-photon background
in Ref.~\cite{Aad:2009wy}.\footnote{Our very approximate background
estimate is $\frac{d\sigma}{dm_{\gamma\gamma}} \simeq
(100\,fb/\gev)\,\lrf{m_{\gamma\gamma}}{150\,\gev}^{-3.7}$
based on Fig.~6 of the $h^0\to \gamma\gamma$ analysis of
Ref.~\cite{Aad:2009wy}.} We find that the statistical signal significance
$S/\sqrt{B}$ is optimized if we require
$60\,\gev < m_{\gamma\gamma} < 90\,\gev$, yielding about
224 total signal events and 44000 background events, corresponding
to $S/\sqrt{B} \simeq 1.1$ with $1\,fb^{-1}$ of data.  Scaling up
by luminosity, a discovery in this inclusive channel based on statistics alone
would require at least $20\,fb^{-1}$ of data.  This number will be strongly
degraded by pile-up and systematics given the very low value of $S/B$,
but it should also be improvable with a more clever analysis,
possibly using a mild $\met$ cut. At this point let us also mention
that we have focused on ATLAS searches here because the CMS di-photon
search has a slightly harder photon $p_T$ requirement of $p_T^{\gamma} >
40\,\gev,\,35\,\gev$~\cite{cms-tdr} which significantly reduces
the signal acceptance.

  Both ATLAS and CMS also plan to perform di-photon Higgs searches
in the exclusive vector boson fusion~(VBF) and associated
$W/h$ and $Z/h$ channels.  These modes have lower production cross-sections,
but the extra products in the events, a pair of forward tagging jets or a
lepton, allow for a significant reduction in background.
To illustrate the potential power of these channels,
we focus on the proposed CMS search for di-photons from $W/h$ and $Z/h$
associated production~\cite{cms-tdr,cms-wzhnote}.  Their strategy is to look
for a photon pair together with at least one lepton ($e$ or $\mu$).
Both photons are required to have $p_T^{\gamma} > 35,\,20\,\gev$,
and $|\eta| < 2.5$.  As above, we deduce a reduced efficiency
$\tilde{\varepsilon}$ for the subset of events with photons passing
these cuts by simulating direct Higgs boson decays to di-photons in PYTHIA.
We obtain $\tilde{\varepsilon} = 0.56,\,0.33$ for $W/h$ and $Z/h$,
respectively, along with a $K$-factor of $1.3$ for the production
cross-section to match onto higher-order
estimates~\cite{Ciccolini:2003jy,Brein:2003wg}.
Combining these factors, we estimate a total of $6.7$ signal
events per $fb^{-1}$ of data in the di-photon invariant mass window
of $20\,\gev < m_{\gamma\gamma} < 90\,\gev$.  The estimated background
rates listed in Tables~10.26 and 10.27 of Ref.~\cite{cms-tdr},
without restricting $m_{\gamma\gamma}$, is about $28\,fb$,
yielding a statistical significance of
$S/\sqrt{B} \simeq 1.26$ with $1\,fb^{-1}$ of data.
This corresponds to a $5\,\sigma$ discovery with about $16\,fb^{-1}$ of data,
and the prospects will improve even further once the backgrounds
are restricted to lie within the signal acceptance window.  Note that this is
considerably better than the estimated significance for the direct di-photon
channel~\cite{cms-tdr} due to the larger signal rate.  More generally,
we expect that any search channel that uses relatively weak
photon $p_T^{\gamma}$ requirements along with an additional handle
to remove backgrounds will be effective for searching for di-photons from
$h^0\to\chi_1^0\chi_1^0$.  Other examples include VBF and $t\bar{t}h$
associated production.

  One additional strategy that could prove useful is to make use
the ability of ATLAS to reconstruct photon ``tracks''.
Using the ECAL alone, the primary di-photon vertex can be identified
to within about $1.5\,cm$ along the beam axis~\cite{Aad:2009wy}.
However, a significant fraction of hard photons interact with material
in the trackers and \emph{convert} to an $e^+e^-$ pair. A converted photon can
therefore give rise to charged tracks allowing for a very precise
determination of the primary photon vertex position. Estimates in
Ref.~\cite{Aad:2009wy} suggest this determination could be as good
as $5\times 10^{-3}\,cm$.  This is useful in regular Higgs
searches for determining the location of primary interaction vertex.
In the case of neutralino decays to gravitinos and photons in
low-scale gauge mediation, vertexing the products of double photon conversions
might allow for a measurement of the neutralino lifetime, even when
the neutralino decays are otherwise ``prompt''.

\section{Conclusions\label{concl}}

  Decays of a SM-like Higgs boson to a pair of neutralinos, each of which
subsequently decays to a photon and a gravitino, can arise in generalized
gauge-mediated supersymmetric scenarios.  This channel leads to a signal
consisting of a di-photon pair plus missing energy at high energy colliders.
In the present work we have investigated this possibility within the
MSSM, and we find that neutralino branching fractions of the Higgs
as large as $0.15$ can be consistent with existing collider bounds.
For branching fractions not too far below this upper bound, both the
Tevatron and the LHC can potentially observe the Higgs boson
through this non-standard decay channel.

  The most promising Tevatron searches for this Higgs mode look
for di-photon events with a small amount of missing energy,
much like in the existing $\dzero$ searches for GMSB.  Our preliminary
analysis suggests that evidence for the Higgs boson can be obtained
with $10\,fb^{-1}$ of data.  Improvements in detection efficiencies
and optimized analysis techniques making use of shape information
in the $m_{\gamma\gamma}-\met$ plane could potentially lead to a discovery,
although a more careful analysis by the collaborations is needed
before a firm conclusion can be drawn.

  At the LHC, this unusual Higgs boson decay channel can be probed
efficiently in associated production channels that use relatively mild
cuts on the photon momenta.  Inclusive production with a mild $\met$ cut
may also be useful for probing this decay mode, though in this case a
more careful study of the backgrounds is required. In principle,
the neutralino mass can be determined from the endpoint of the di-photon
mass distribution in this channel once the Higgs boson mass is known.
Photon conversions may also provide information about the lifetime
of the unstable neutralino and the mass of the gravitino.

  In the present work we have focused on Higgs decays to neutralinos
in the MSSM, but even more possibilities can arise in extensions with
additional singlets~\cite{nmssm,n2mssm}.  In this case the Higgs boson
can potentially have significant branching ratios to any neutralinos that
are mostly bino or singlino.  Depending on the precise spectrum and
pattern of mixing angles, various signatures involving photons
and missing energy are possible.  Another interesting possibility occurs
if the spectrum also contains a light pseudoscalar $a_1$,
allowing for example the decays $h^0 \to \chi_1^0\chi_1^0$
with $\chi_1^0 \to a_1\tilde{G}$.  This could lead to collider signatures
similar to those considered in Ref.~\cite{Dermisek:2005ar},
but with modified kinematics and missing energy.  We hope to study these
possibilities in a future work~\cite{future}.

  Finally, we would like to emphasize the importance of searching broadly
for new physics in upcoming collider data.  In the present analysis we
have found that in many cases cuts imposed while optimizing searches
for new physics in other channels also cut out the majority of the
signal in the non-standard Higgs channel we are interested in.
Given the wide variety of possibilities for new physics beyond the standard
model, we feel that it is most prudent to cast our experimental nets
as widely as practicable.

\section*{Acknowledgements}

We thank Spencer Chang, Mark Cooke, Lorenzo D\'\i az-Cruz, Thomas Gadfort,
Yuri Gershtein, Steve Godfrey, Jack Gunion, Graham Kribs, Heather Logan,
Shulamit Moed, John Ng, Aaron Pierce, Matt Schwartz, Carlos Wagner,
and James Wells for helpful comments and conversations.
We are especially indebted to Yuri Gershtein for answering our many
questions about photon identification at $\dzero$ and Aaron Pierce
for comments on the manuscript.
This work is supported in part by the Harvard Center for the Fundamental
Laws of Nature, by NSF grant PHY-0556111~(DP), by the DOE~(JM and DM),
and by NSERC~(DM).  DM would also like to
thank the \'Ecole de Physique des Houches and the Aspen Center for
Physics where portions of this work were completed.  DP would like to thank
the Galileo Galilei Institute for its hospitality during the completion of this
work.



\end{document}